\documentstyle[galley,epsf]{mn}
\newif\ifAMStwofonts

\ifoldfss
  \ifCUPmtlplainloaded \else
    \NewTextAlphabet{textbfit} {cmbxti10} {}
    \NewTextAlphabet{textbfss} {cmssbx10} {}
    \NewMathAlphabet{mathbfit} {cmbxti10} {} 
    \NewMathAlphabet{mathbfss} {cmssbx10} {} 
  \fi
  \ifAMStwofonts
    \ifCUPmtlplainloaded \else
      \NewSymbolFont{upmath} {eurm10}
      \NewSymbolFont{AMSa} {msam10}
      \NewMathSymbol{\upi}     {0}{upmath}{19}
      \NewMathSymbol{\umu}     {0}{upmath}{16}
      \NewMathSymbol{\upartial}{0}{upmath}{40}
      \NewMathSymbol{\leqslant}{3}{AMSa}{36}
      \NewMathSymbol{\geqslant}{3}{AMSa}{3E}

    \fi
  \fi
\fi 

\ifnfssone
  \newmathalphabet{\mathit}
  \addtoversion{normal}{\mathit}{cmr}{m}{it}
  \addtoversion{bold}{\mathit}{cmr}{bx}{it}
  \newmathalphabet{\mathbfit} 
  \addtoversion{normal}{\mathbfit}{cmr}{bx}{it}
  \addtoversion{bold}{\mathbfit}{cmr}{bx}{it}
  \newmathalphabet{\mathbfss} 
  \addtoversion{normal}{\mathbfss}{cmss}{bx}{n}
  \addtoversion{bold}{\mathbfss}{cmss}{bx}{n}
  \ifAMStwofonts
    \ifCUPmtlplainloaded \else
      \UseAMStwoboldmath
      \makeatletter
      \new@mathgroup\upmath@group
      \define@mathgroup\mv@normal\upmath@group{eur}{m}{n}
      \define@mathgroup\mv@bold\upmath@group{eur}{b}{n}
      \edef\UPM{\hexnumber\upmath@group}
      \new@mathgroup\amsa@group
      \define@mathgroup\mv@normal\amsa@group{msa}{m}{n}
      \define@mathgroup\mv@bold\amsa@group{msa}{m}{n}
      \edef\AMSa{\hexnumber\amsa@group}
      \makeatother
      \mathchardef\upi="0\UPM19
      \mathchardef\umu="0\UPM16
      \mathchardef\upartial="0\UPM40
      \mathchardef\leqslant="3\AMSa36
      \mathchardef\geqslant="3\AMSa3E
    \fi
  \fi
\fi 

\ifnfsstwo
  \DeclareMathAlphabet{\mathbfit}{OT1}{cmr}{bx}{it}
  \SetMathAlphabet\mathbfit{bold}{OT1}{cmr}{bx}{it}
  \DeclareMathAlphabet{\mathbfss}{OT1}{cmss}{bx}{n}
  \SetMathAlphabet\mathbfss{bold}{OT1}{cmss}{bx}{n}
  \ifAMStwofonts
    \ifCUPmtlplainloaded \else
      \DeclareSymbolFont{UPM}{U}{eur}{m}{n}
      \SetSymbolFont{UPM}{bold}{U}{eur}{b}{n}
      \DeclareSymbolFont{AMSa}{U}{msa}{m}{n}
      \DeclareMathSymbol{\upi}{0}{UPM}{"19}
      \DeclareMathSymbol{\umu}{0}{UPM}{"16}
      \DeclareMathSymbol{\upartial}{0}{UPM}{"40}
      \DeclareMathSymbol{\leqslant}{3}{AMSa}{"36}
      \DeclareMathSymbol{\geqslant}{3}{AMSa}{"3E}
    \fi
  \fi
\fi 

\ifCUPmtlplainloaded \else
  \ifAMStwofonts \else 
    \def\upi{\pi}
    \def\umu{\mu}
    \def\upartial{\partial}
  \fi
\fi

\begin{document}
\title{R\,Coronae Borealis Stars at Minimum Light -- UW Cen}
\author[N. Kameswara Rao,  Bacham E. Reddy \& David L. Lambert]
       {N. Kameswara Rao$^1$, Bacham E. Reddy$^1$, David L. Lambert$^2$\\
       $^1$Indian Institute of Astrophysics, Bangalore 560034, India\\
       $^2$The W.J. McDonald Observatory, University of Texas, Austin, TX 78712-1083, USA\\}
\date{Accepted .
      Received ;
      in original form  }

\pagerange{\pageref{firstpage}--\pageref{lastpage}}
\pubyear{}

\maketitle

\label{firstpage}

\begin{abstract}

Two high-resolution optical spectra of the R Coronae Borealis star
UW Cen in decline are discussed. 
A spectrum from mid-1992 when the star had faded by three magnitudes
shows just a few differences with the spectrum at maximum
light. The ubiquitous sharp emission lines seen in R CrB at a
similar drop below maximum light are absent. In contrast, a spectrum
from mid-2002 when the star was five magnitudes below maximum
light shows an array of sharp emission lines and a collection
of broad emission lines. Comparisons are made with spectra of R CrB
obtained during the deep 1995-1996 minimum. The many common
features are discussed in terms of a torus-jet geometry.

\end{abstract}


\section{Introduction}

R Coronae Borealis stars (here, RCBs) are H-poor  supergiants that decline
in brightness unpredictably and rapidly  by up to  8 magnitudes
to remain at or near minimum light for several weeks to
months, and even years in some cases.
It is generally accepted that the declines 
are due to the  formation of a cloud of carbon soot that obscures the
stellar photosphere (O'Keefe 1939). At present, spectroscopic coverage
of RCBs in decline is quite limited but common signatures
seem  to exist.

In decline, two kinds of emission lines dominate the optical spectrum of a
RCB (Herbig 1949; Payne-Gaposchkin 1963)):
 a rich set of sharp lines (FWHM $\sim 12$
km s$^{-1}$), and a sparse set of broad lines (FWHM $\sim 300$ km s$^{-1}$).
Sharp lines first appear very early in a decline
and persist through deep minima into the recovery to maximum light.
Initially and for a few days, the emission lines are dominated by high-excitation
transitions of abundant species (e.g., C\,{\sc i} and O\,{\sc i} - Rao
et al. 1999). After a few days and through the minimum to recovery,
the  sharp lines are
of low excitation with singly-ionized metals
well represented. Broad lines, which are seen only when the star has
faded by several magnitudes and do not have a common profile,
 may include the following
carriers: He\,{\sc i} triplet series,
Na\,{\sc i} D, and [N\,{\sc ii}] lines, i.e., a mix of high and
low excitation lines. Broad blue-shifted absorption lines
have been seen to accompany Na\,D, Ca\,{\sc ii} H and K,
and the He\,{\sc i} lines at 10830 and
3889 \AA, especially at and following minimum light. 
 In deep minima, the photospheric absorption
lines are `veiled', i.e., the lines become
very shallow and broad.

These spectral features may be common to all RCBs. 
Their discovery and
definition rested largely on   well observed declines of the two
brightest RCBs  -- R CrB  (Payne-Gaposchkin 1963) and RY Sgr
(Alexander et al. 1972).  
Evidence that the features may be ubiquitous is found in the
valuable contributions to the spectroscopic literature of RCBs 
observed through a decline  by Cottrell, Lawson, \& Buchorn
(1990) for R CrB, Skuljan \& Cottrell (1999) for S Aps and RZ Nor,
and Skuljan \& Cottrell (2002a) for V854 Cen.
Observations of a few stars in deep declines have also shown that
a RCB in such a decline has common spectroscopic features, as we indicate in
Section 7.
 Our reference point
will be the extensive coverage at high-spectral resolution of R CrB
in the 1995-1996 decline (Rao et al. 1999). 

In this paper, we discuss  high-resolution spectra of the southern
RCB UW Cen taken  in mid-1992 and mid-2002 when the star
was in a deep decline. We comment also on low resolution spectra of UW Cen 
obtained prior to 2002 when UW Cen was in the same
long decline
(Skuljan \& Cottrell 2002b)
 UW Cen with R CrB and RY Sgr is a 
 `majority' member of the RCB class in the terminology
introduced by Lambert  \& Rao (1994). UW Cen's effective temperature
($T_{\rm eff}$ = 7500 K) and
surface gravity ($\log g = 1.0$) are representative of the majority
stars (Asplund et al. 2000).

UW Cen is one of the very  few RCBs with a visible circumstellar reflection
nebula (Pollacco et al. 1991).
The nebula varies in appearance, possibly due to the
formation and dissipation of dust clouds affecting the
illumination of the nebula
(Clayton et al. 1999). At maximum light, UW Cen shows  
light  (amplitude
 less than 0.2 mag in V)  and radial velocity variations 
typical of RCBs
 (Lawson \& Cottrell 1997).

\section{Observations}

High-resolution optical spectra were obtained in 1991, 1992, and 2002 
 at CTIO, using the 4 meter
Blanco telescope and Cassegrain echelle spectrometer.
 The star was
at maximum light in 1991, but in decline
in 1992 and 2002.
The  1991 July 15 (JD 2448453)  spectrum 
covered the spectral range 5540 to 6780~\AA\ at a resolving power of 12,500.
A spectrum
obtained on 1992 May 22
 (JD 2448764)   at a resolving power of
30,000  covered 
the spectral region from 5480 to 7090 ~\AA.
The AAVSO database of variable star observations
(http://www.aavso.org/data/lcg/) shows that the star
was observed in 1992 at about 12th magnitude, or about
3 magnitudes below maximum light. 
In this decline,
which began 90 days before our observation, the star faded to 15th
- 16th magnitude by about JD 2448730  before brightening slightly to 12th
magnitude and then fading again.  
 Maximum light was not attained until about one
year after our observation. 

Two 30 minute exposures of  UW Cen were obtained on 2002 June 20 (JD 2452446.54).
 The spectrograph was  set to record the wavelength interval
4900 to 8250~\AA\ in 45 orders. Spectral coverage was
complete between these limits. 
A resolving power of
35,000  was achieved. 
The light curve from the AAVSO database
shows that the star went into decline in late-1998.
As typical of RCBs, the decline was rapid. The star was fainter than
 V $\simeq$ 13 from mid-1997 (V = 9.1 at maximum), and, in particular,
at minimum light with V $\simeq 15$ or possibly fainter from
early-1999 to the beginning of 2001 when it brightened gradually reaching 13th
 magnitude by early  2002.
At the time of our observation, UW Cen had faded again and
was at V $\simeq 14$ or
about 5 magnitudes below maximum brightness. 
 In the following
months and through the 2003 observing season, it remained very
faint.

\begin{figure}
\epsfxsize=8truecm
\epsffile{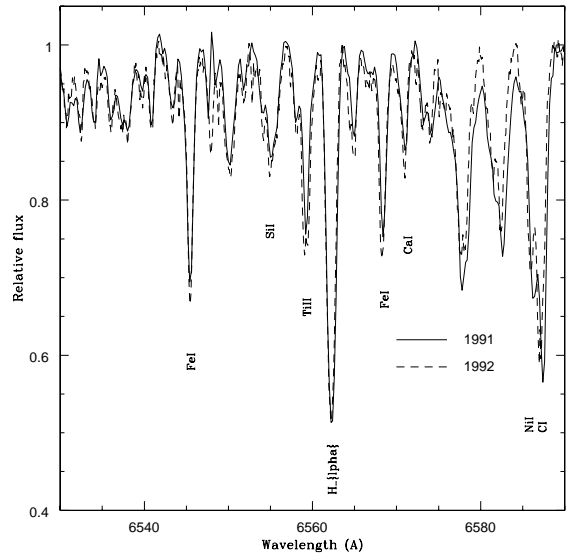}
\caption{Comparison of UW Cen spectra obtained in 1991(light maximum) and 1992 (light minimum).
 Note the good match of the lines including H$\alpha$.}
\end{figure}

\section{UW Cen's 1991 spectrum}

The maximum-light spectrum obtained in 1991 
 shows a good match to the 1992  spectrum when the star was
about three magnitudes below maximum light.
 The  photospheric
 absorption lines are broad enough not to be affected by the difference
 in resolving power of the spectra.
Figure 1 shows the superposition of the 1991 (full line) and 1992
(dashed line) spectra in the region of H$\alpha$. It is clear that
a majority of
the lines  have similar
profiles in 1991 and 1992.
The 1992 spectrum was used by 
Asplund et al (2000) for their abundance analysis.
Analysis of the 1991 spectrum would confirm that analysis.

The radial velocity in 1991 from 32 unblended lines 
across the interval 5800 - 6750 \AA\ is -31.8$\pm$2.6 km s$^{-1}$
with two strong Ba\,{\sc ii} lines indicating a velocity of about
-20 km s$^{-1}$.
Lawson \& Cottrell (1997) obtained radial velocities on 10 occasions
between 1991 May and 1991 September from spectra of the 6540 - 6640 \AA\ interval
with a resolving power of 10,000. Their two observations  close to the
time of our observation give a velocity of about -21 km s$^{-1}$, a value
differing by 12 km s$^{-1}$ from our measurement.


\begin{figure}
\epsfxsize=8truecm
\epsffile{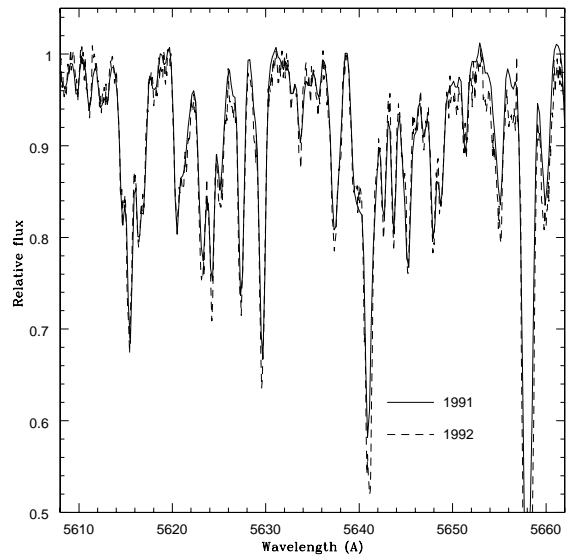}
\caption{Comparison of spectra of UW Cen obtained in 1991 and 1992.
 Note the strengthening of the  Ba\,{\sc ii}  6141 \AA\ line in the 1992 spectrum.}
\end{figure}

\section{UW Cen's  1992 Spectrum}

When observed in 1992, UW Cen was about 3 magnitudes below
maximum light yet the 1992 spectrum closely resembles the
1991 maximum light spectrum. The prominent difference is that
certain strong low excitation  lines 
are strengthened in the 1992 spectrum, notably 
strong lines of Sc\,{\sc ii} and Ba\,{\sc ii}:
 Figure 2 
 shows the interval 6095 --6160 \AA\ where strengthening of
the Ba\,{\sc ii} 6140 \AA\ line in 1992 is conspicuous. Strengthened lines
 are slightly red-shifted ( 7 km s$^{-1}$) relative to the photospheric velocity of
\ $-43.8\pm 1.5$ km s$^{-1}$, as measured from 23 lines.

\begin{figure}
\epsfxsize=8truecm
\epsffile{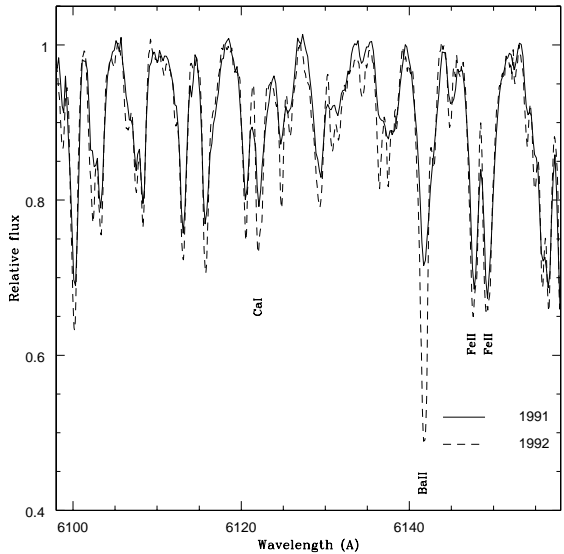}
\caption{Comparison of UW Cen spectra around the  Na\,{\sc i}  D lines
 for the 1992 and 2002 minima.
 Note the presence of sharp emission  in the core of a  D line 
in the 1992 (indicated by arrows) at the same
velocity as the prominent emission in the 2002 spectrum.
 Blue-shifted shell absorption components
 are present in the  1992 spectrum at -157 and -194 km s$^{-1}$.}
\end{figure}
 
It is a well known fact that when
the well observed stars R CrB and RY Sgr decline, sharp emission lines
appear in great numbers across the optical spectrum from
 very early in the decline and  until late in the
recovery phase. 
The striking aspect of UW Cen's 1992 spectrum is that the only  sharp 
emission lines are 
at the Na D lines.
 Figure 3 shows  sharp emission cores in the
 Na  D lines  
at a radial velocity of -35.0 km s$^{-1}$. This emission is 
red-shifted by about 9 km s$^{-1}$ with respect to the photospheric
absorption lines.

Figure 3 also shows blue--shifted  absorption Na D lines in the 1992 spectrum
with absorption minima
at heliocentric radial velocities of $-$157 and $-$194 km s$^{-1}$.
The high-velocity Na D components are attributed to a shell
ejected and accelerated earlier.
In analogy with the 1995-6 minimum of R CrB (Rao et al. 1999),
 it would be expected that the  shells
were  ejected at about the time that minimum light
was attained, some 60 to 90 days before our spectrum was taken,
and accelerated to the observed velocities.


\section{UW Cen's  2002 Spectrum}

The 2002 spectrum  resembles  that of R CrB observed 
 at  minimum light
(Rao et al. 1999): sharp emission lines of singly-ionized
metals are prominent; broad emission lines of Na D and He\,{\sc i} 7065 \AA\
stand out and others are revealed by close inspection; the continuum is
almost unmarked by the photospheric absorption lines prominent in the spectrum
at maximum light. Detailed discussion of these components follows.

\subsection{The sharp emission lines}

Permitted and forbidden atomic lines are present as sharp emission lines.
The list of atomic lines is dominated by low excitation lines of singly 
ionized metals, e.g.,
 Sc\,{\sc ii}  (RMT 19, 28, 29, 31, 26, 27),
 Ti\,{\sc ii} (RMT 101, 92, 91, 69, 103,
70, 86, 71), Fe\,{\sc ii} (RMT 74, 40, 48, 49, 42), Y\,{\sc ii} (RMT 26, 20),
 and Ba\,{\sc ii}  (RMT 1, 2). Except for the Na I D  lines and K I (RMT 1),
sharp emission lines  of neutral atoms are not present.
 Li\,{\sc i} at 6707 \AA, a strong absorption line at maximum light, is
not present either in absorption or in emission.
Molecular emissions (C$_2$ and CN) are not present.
 The sharp lines are  slightly broader than the instrumental width
 (as assessed from the [O\,{\sc i}] night sky 
emissions):  the sharp lines' FWHM,  after correcting for the instrumental
width is 11.4 km s$^{-1}$.
     The radial velocity as determined from 53 sharp lines
 is -34.6$\pm$1.9 km s$^{-1}$.
Our spectrum was not flux-calibrated, but the similarity
between  UW Cen's and R CrB's  lines suggests that the excitation
temperature of UW Cen's emitting region is similar to that
derived for R CrB, i.e., $T_{\rm exc} \simeq$ 4000 K.

\begin{figure}
\epsfxsize=8truecm
\epsffile{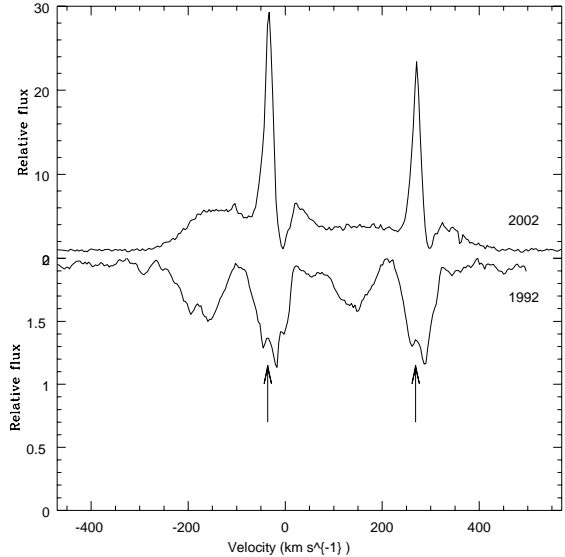}
\caption{Profiles of sharp emission lines showing red-shifted absorption. 
The peak radial velocity of the emission is at  -34 km s$^{-1}$
and the weak absorption component is at about 6 km s$^{-1}$.}
\end{figure}

     Some  strong emission lines are accompanied
 by red-shifted absorption (see Figure 4).
The absorption appears at velocities between
-4 to -11 km s$^{-1}$.
  It is unlikely that the absorption component
is simply the result of  emission superposed on the photospheric line. Since the
absorption-free unblended sharp emission lines are symmetrical about the
emission peak, it is unlikely that the absorption component has
eaten into the sharp emission line.
    
A few forbidden transitions are seen as sharp emission lines:
[Ca\,{\sc ii}] at   7291 \AA\ and  7323 \AA,  
and [Fe\,{\sc ii}] at  7155 \AA.
 The sharp [Ca\,{\sc ii}] lines are superposed on  broader
components (Figure 5).
 The sharp forbidden components are at a velocity of -33 km s$^{-1}$, a value
consistent with  sharp permitted emissions. The line width (FWHM),
after correcting for the  instrumental width is about 13 km s$^{-1}$,
also consistent with other sharp emission lines. 
 
\begin{figure}
\epsfxsize=8truecm
\epsffile{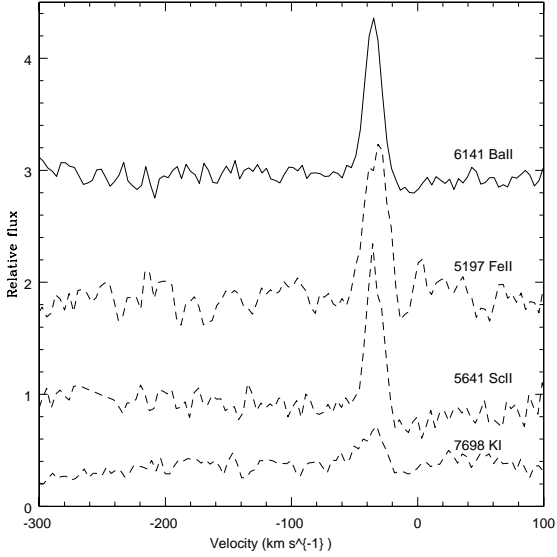}
\caption{The  [Ca\,{\sc ii}] lines in the 2002 minimum spectrum of UW Cen.}
\end{figure}

\subsection{\bf The broad emission lines}

Broad emission lines include 
He\,{\sc i} at  5876 \AA, and 7065 \AA, the Na  D lines, and lines of 
 [N\,{\sc ii}],  [S\,{\sc ii}], and [Ca\,{\sc ii}].
The He\,{\sc i} lines have a roughly parabolic line
profile.
The [N\,{\sc ii}] profiles are double-peaked, but the weaker
[S\,{\sc ii}] lines are single-peaked.

\begin{figure}
\epsfxsize=8truecm
\epsffile{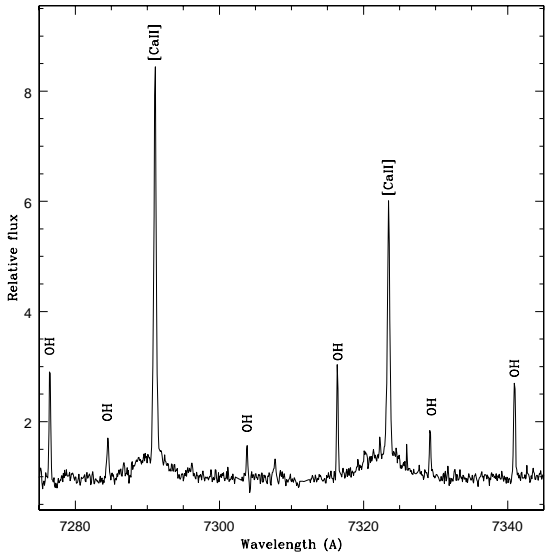}
\caption{He\,{\sc i} lines at 5876\AA\ and 7065\AA\ in the
 2002 minimum spectrum. 
 In contrast to the parabolic He\,{\sc i} profiles,
the  [N\,{\sc ii}] line shows a double-peaked emission profile.}
\end{figure}

\subsubsection{He\,{\sc i} lines}

The He\,{\sc i} lines at 5876\AA\ and 7065\AA\ are  seen as broad
emission lines with a P- Cygni profile (Figure 6).
The rest wavelengths  (5875.652 \AA,
and 7065.277 \AA) are taken from laboratory wavelengths with the unresolved
components weighted by their theoretical intensities for optically
thin emission.
The 7065 \AA\ line may have a P Cygni profile with
weak absorption  to -460 km s$^{-1}$  and  maximum  absorption
at  -305 km s$^{-1}$. The broad emission is centered 
at -66 km s$^{-1}$, extending on the redside to about 112 km s$^{-1}$.
 It also may be noted that there is no sharp emission superposed on these
broad lines.

\subsubsection{Na  D lines}

The Na  D lines show a broad emission profile with
 superposed sharp emission (Figure 3).
The blue edge of broad emission of D2 extends to -273 km s$^{-1}$. 
 The red edge of D1 extends to 
$+$128 km s$^{-1}$.
The profile (after removing the sharp emission) can be  fit by a double
peaked profile consisting of two Gaussians, each with
a FWHM of 127 km s$^{-1}$, a separation of  144.0 km s$^{-1}$,
and of roughly equal flux.
The flux ratio of the D2 to D1 broad emission lines is about
1.8, close to ratio of 2.0 for optically thin emission lines.

\begin{figure}
\epsfxsize=8truecm
\epsffile{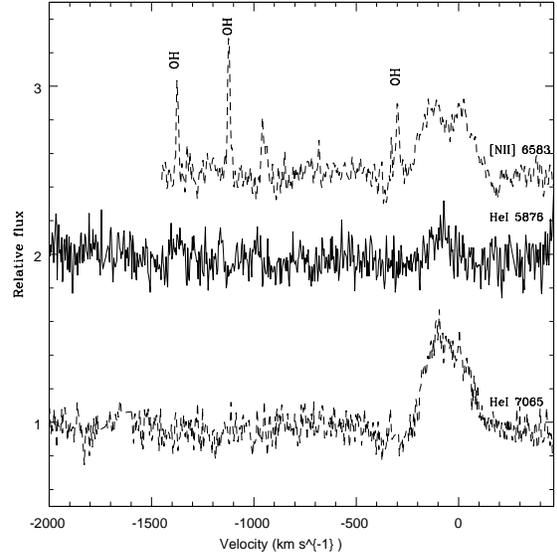}
\caption{Lines of [N\,{\sc ii}] and [S\,{\sc ii}] in the spectrum of
the  2002 minimum of  UW Cen.}
\end{figure}

\subsubsection{Forbidden lines}

[Ca\,{\sc ii}]:  The Ca$^+$ ion is expected to contribute broad emission lines
at   3968 \AA\ and 3933 \AA\ (the 
H and K lines), the infrared triplet lines at 8542 \AA, 8662 \AA, and 8498 \AA,
and the forbidden lines at 7291 \AA\ and 7323 \AA. Our spectrum
does not cover the wavelength regions of the permitted lines.
The broad component of the forbidden lines is
of approximately the same strength for the two lines (Figure 5).
 The velocity range at the base of the broad emission extends from
about $-250$ km s$^{-1}$ to $+103$ km s$^{-1}$  with a
centre at about $-61$ km s$^{-1}$, and FWHM of about 222 km s$^{-1}$.

[N\,{\sc ii}]: The lines  6583 \AA\ and 6548 \AA\ lines
 are  present as
broad emissions. The 5754 \AA\ line  is not detected.
Figure 7 shows  the 6583 \AA\ [N\,{\sc ii}] line  as double-peaked.
Rest wavelengths
 of  6583.454\AA\ and 6548.08\AA\ have
been adopted (Dopita et al. 1997).
 The velocity range at the base extends from $-245$ km s$^{-1}$
 to $+135$ km s$^{-1}$. The red and blue peak
are separated by 145 km s$^{-1}$ with the blue peak at  $-128$ km s$^{-1}$ 
and the red peak at $+17$ km s$^{-1}$ for the  6583 \AA\ line, the stronger line
of the pair. The peaks are  about
equal strength, and of similar width (FWHM of 120 km s$^{-1}$).
The mean velocity is about $-56$ km s$^{-1}$.

[S\,{\sc ii}] : The red lines at 6717 \AA\ and  6731 \AA\ are present in
emission. Rest wavelengths are  adopted from Dopita et al, (1997),
namely,
6716.472 and 6730.841 \AA. The [S\,{\sc ii}]
 line  profiles  differ in  a surprising
way from the 
profiles of the [N\,{\sc ii}] lines: both [S\,{\sc ii}] lines
show the red but not the blue peak of the [N\,{\sc ii}]
profiles (Figure 7).
The  red peaks of the [S\,{\sc ii}] and [N{\sc ii}] lines 
have approximately the same width.
The equivalent width of the  6731 \AA\ line is about twice
that of the   6717 \AA\ line. This ratio for gas at a temperature
 of about 10000 K (or less)
corresponds to an electron density $n_e \simeq 10^{3.7}$
cm$^{-3}$ (Aller 1984).

\subsection{Absorption lines}

The normal photospheric spectrum is absent.
Figure 8  shows the region of the strong photospheric
line of Si\,{\sc ii} at 6347 \AA\  
from the 2002 and 1992 spectra.  Clearly, photospheric
lines are `veiled' in the 2002 spectrum. 
 The radial velocity estimated from some 15 lines is -4.3$\pm$5.7 km s$^{-1}$ where
the large standard deviation reflects the difficulty of measuring these
weak broad lines.
 
\begin{figure}
\epsfxsize=8truecm
\epsffile{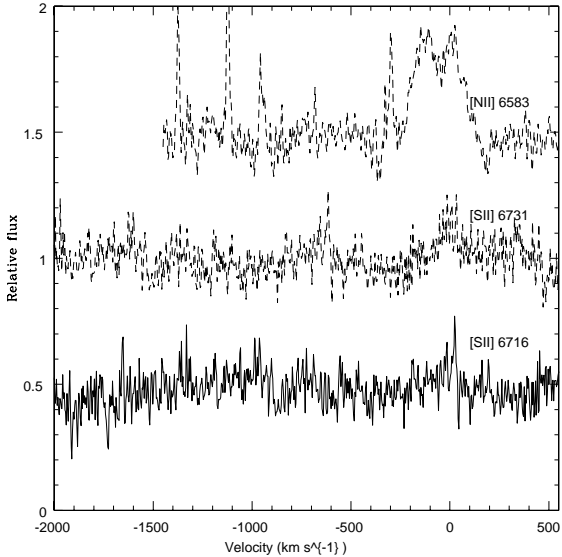}
\caption{Absorption lines in spectra of UW Cen at the 2002 and 1992 minima.
Note the shallow Si\,{\sc ii} 6347 \AA\ line in the 2002 minimum spectrum}
\end{figure}

A few   narrow absorption lines do appear in the 2002 spectrum 
but these are not
photospheric lines.
Figure 9 shows several low excitation lines of Fe\,{\sc i}  
in absorption without any accompanying emission. They provide a
radial velocity of -9.7$\pm4.3$ km s$^{-1}$ from six lines, a velocity
similar to that of the red-shifted absorption components
associated with the sharp emission 
lines. 

\begin{figure}
\epsfxsize=8truecm
\epsffile{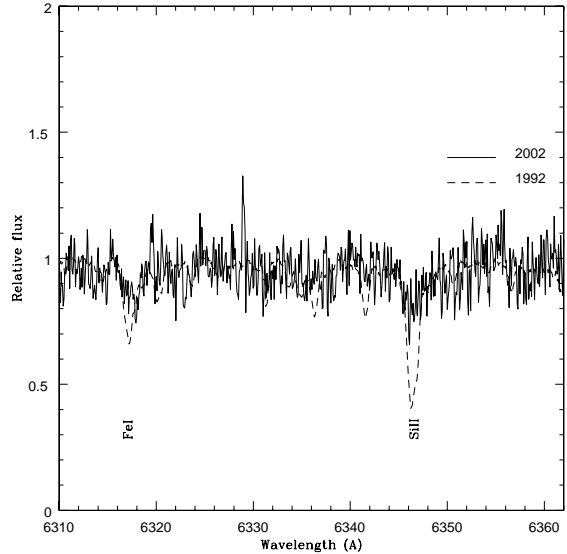}
\caption{Low excitation Fe\,{\sc i} transitions in the 2002 minimum
spectrum showing a sharp absorption line unaccompanied by emission.}
\end{figure}

\section{UW Cen  and R CrB in decline}

The components of UW Cen's spectrum in the 2002 deep minimum are
similar to those seen
during the  1995 - 96 minimum of R CrB 
(Rao et al. 1999).
 These
components include the veiled  photospheric spectrum, low
excitation  sharp emission
lines,  broad emission
lines, and  high velocity blue-shifted absorption lines seen in the
Na D lines. 
Other minima of R CrB show the same components,
as do the limited observations of other R CrB stars in decline.
In this section, we compare spectroscopic properties of 
UW Cen and R CrB at minimum light.

\subsection{Sharp emission lines}

Sharp  high excitation emission
lines were seen in R CrB for a short period beginning just after the
onset of the decline. Since our UW Cen spectrum was taken long after the
decline's onset it is not suprising that the equivalent lines are
absent.

Lower excitation lines of singly-ionized metals and neutral atoms
 are seen in R CrB following disappearance of the
high excitation lines.
 These lines  show a similar level of 
excitation for the two stars. 
The line widths are also similar for the two stars.
The sharp emissions made their initial appearance in the cores of 
R CrB's absorption lines  
when the star had faded by  about  3 magnitudes.
At the time of our 1992  observation, when UW Cen was
about 2.7 magnitudes fainter than at maximum,
sharp emission lines  are detected only in the  Na  D line cores,
suggesting that the emitting region was weaker or more
obscured than in the case of R CrB.
 Sharp emission lines are prominent in the
2002 spectrum when UW Cen was about five magnitudes below maximum light. 

That the sharp Na  D line emission for UW Cen
occurs at the same velocity ($-35$ km s$^{-1}$) in 1992 and 
2002 minimum suggests that 
the emitting gas may be a semi-permanent feature detached  
from the stellar photosphere. This is consistent with observations of
R CrB showing a constant velocity for sharp emission lines and
a variable velocity for photospheric lines.

Red-shifted absorption  associated with sharp emission lines
is seen in UW Cen and was seen
in the early stages of the R CrB decline, and also in a decline of
V854 Cen (Rao \& Lambert 2000). 
UW Cen's  2002 spectrum also shows this absorption  component in low excitation
Fe\,{\sc i} lines without accompanying emission. 
Comparison of the  1991 maximum and 1992 minimum spectra
shows 
strong absorption lines of Ba\,{\sc ii} redshifted relative to
 other absorption lines.
The red-shifted absorption suggests that cool gas is falling toward
the star -- from the detached region responsible for the
sharp emission lines?

\subsection{Broad emission lines}

The inventory of broad emission lines in UW Cen in 2002 does not
precisely  duplicate that for R CrB at its minimum across
the common wavelength interval.
The  profile of a line in UW Cen and the same
line in R CrB  are generally similar, most notably  the He\,{\sc i} lines
have a quasi-parabolic profile, and the [N\,{\sc ii}] lines
are double-peaked. 
The [S\,{\sc ii}] lines are seen in UW Cen (also, V854 Cen, Rao \& Lambert
1993) but not in R CrB. The K\,{\sc i} 7664 \AA\ and 7699 \AA\ lines
were seen in R CrB but not in UW Cen. The C$_2$ Swan bands were in
emission in R CrB but not in UW Cen. These differences may reflect
differences in composition and physical conditions
rather than major structural differences.

The low excitation sharp emission lines in R CrB, UW Cen, and all other
stars where they have been seen occur with a small blueshift
relative to the photosphere. 
Broad emission lines   share neither the velocity of the
sharp lines nor the  velocity of the photosphere, and appear
not to  be constant for a given
star. 
For UW Cen in 2002, 
the broad emissions are displaced to the blue by 26 km s$^{-1}$ from the
sharp emission lines (mean broad line velocity is $-61$ km s$^{-1}$).
This mean velocity is outside the range of
reported photospheric velocities:  -10.1 to -26.8
km s$^{-1}$ (Lawson and Cottrell 1997) , -29.3 km s$^{-1}$ (Herbig 1994,
private communication)
and -43.8 km s$^{-1}$ (this paper).
In the case of R CrB in the 1995-1996 minimum, the broad lines were 
blue shifted by about 20 km s$^{-1}$.
For RY Sgr in separate declines
 Alexander et al. (1972) and Spite \& Spite (1979) reported
differing redshifts of the broad lines relative to the sharp lines, but
Asplund (1995) found broad and sharp lines at the same velocity. 
An especially intriguing star is S Aps where the broad Na D lines
were blue-shifted by about 100 km s$^{-1}$ relative to the
sharp emission lines with their normal small blueshift
with respect to the photosphere (Goswami et al. 1997).

\subsection{Blue-shifted Na D absorption}

Observations of R CrB in the 1995-1996 minimum showed the
blue-shifted high-velocity Na D absorption  developing in
strength and velocity shortly after minimum light. 
Such high-velocity gas has been
reported for other declines of R CrB  and other R CrB stars.
It is then not surprising that the 1992 spectrum of 
UW Cen obtained about two months after the star had reached
minimum light should show high-velocity absorption in the
Na D lines.

High-velocity absorption with a profile like that of
the 1992 components is not present in our 2002 spectrum.
Very broad shallow absorption could be difficult to
impossible to detect given the presence of the broad
emission line.
Absorption at velocities less than about $-$250 km s$^{-1}$
is definitely absent.

Observations of the Na\,D lines are reported
by Skuljan \& Cottrell (2002b) who observed UW Cen at low resolution
($\Delta\lambda \simeq 3 - 6$ \AA)
during a partial recovery to V $\simeq$ 13 -- 14
in the deep extended decline in which our 2002
spectrum was obtained. These Na\,D observations were obtained between
534 and 622 days from the onset of the decline, and about four
years before our observation. 
The spectra appear to show blue-shifted absorption at a velocity
of about -220 to -300 km s$^{-1}$. Such a component is not
present in our spectrum. (Skuljan \& Cottrell did not detect
broad Na\,D emission lines but remark that `they are probably at or
below the noise level in the continuum'.)

\section{DUSTY TORUS AND BIPOLAR JETS}

In line with contemporary fashion for models of mass-losing luminous
low mass stars (Lopez 1999; Rao, Goswami \& Lambert 2002) 
including RCB stars (Rao \& Lambert 1993; Clayton et al. 1997),
 we invoke the geometrical combination of  bipolar jets and an 
equatorial torus of dust and gas.

There are observational hints
that there is
a preferred plane (i.e., a torus)
 for dust formation and ejection (Clayton et al. 1997; Rao \& Raveendran
1993). 
 Observational evidence, especially  measurements of the
infrared flux emitted by the circumstellar dust, suggests
 that the dusty region is comprised
of clouds (Feast 1986; Feast et al. 1997) and a decline results when a
new cloud forms along
the line of sight. If  dust formation is restricted to a preferred
(equatorial) plane and clouds confined
to a torus, dust-forming RCB stars observed
at high inclination to the torus will not exhibit declines but
will  possess an infra-red excess. The stars Y Mus, XX Cam may be such
examples (Clayton 1996; Walker 1986 ) At the other extreme, there may be
stars with the
 torus perpendicular to the plane of the sky and, if
the torus is optically thick, these RCBs will be difficult
to identify.

 We speculate  how a torus-jet  
model may account for the spectroscopic signatures of an RCB in decline,
including the sharp emission lines, the 
broad emission lines, the veiled photospheric
lines, and the high-velocity blue-shifted Na D absorption:
\begin{itemize}
\item
Sharp  low-excitation emission lines, we
suppose, originate from the base of the jets where the jet
velocity is low.
 Depending on
the geometry of the jet and torus,  the viewing angle and
the size of the new cloud,
  the base of the jet
on the  far side 
may be obscured.
In which case, the
sharp lines will always have a net blueshift determined by the jet
outflow velocity near its base, the angle of inclination and the volume of
the emitting region unobscured by dust. The characteristic
velocity of the lines when the star is in decline is
$-$10 km s$^{-1}$ relative to the photosphere. 
\item
Double-peaked broad  emission lines (e.g., [N\,{\sc ii}] but not
He\,{\sc i})  are
identified with regions of the jets quite distant from the
star and visible to us along paths not intersecting the
newly formed cloud and the dusty torus. 
The double-peaked profiles have a ready
explanation: the blue-shifted peak comes from the jet
nearer to us and the red-shifted peak from the opposing jet. The
details of the profile depend on the jet velocity and the opening
angle and the angle of inclination to the line of sight.
The mean velocity could be, as observed for UW Cen, blue-shifted
relative to the photospheric velocity if the receding jet is
partially obscured by the dusty torus. However, this
construction would likely predict unequal fluxes in the
blue and red halves of the line, which is not the case (see also
Rao et al. 1999 for R CrB).

If there is a steep transition from subsonic to supersonic flow (i.e.,
a Parker-type wind), 
lines with  profiles intermediate between the sharp and broad
lines will
be absent, as observed.
This geometry is consistent with Whitney et al.'s (1992) observation that
the Na D broad lines from V854 Cen are unpolarized even though the
local continuum is significantly polarized. It is also consistent
with the indication that the broad line flux  is
probably  constant through a decline, as shown by
Skuljan \& Cottrell (2002a) for V854 Cen, and  by Rao et al. (1999)
for R CrB.

\item
 Judged by differences in profile and
derived physical conditions, the He\,{\sc i} lines cannot
be produced by the gas responsible for the [N\,{\sc ii}] and 
similar lines. 
The He\,{\sc i} emission is here identified with the dusty
equatorial torus.
 Gas in the clouds causing a decline  is accelerated quite
quickly
to high velocities, as revealed by the blue-shifted Na D absorption.
Dust ejection is known to occur in localized regions. 
Then, the torus will be populated by many high velocity clouds.
Collisions in the torus
 between the high velocity atoms in a high-velocity  dust cloud and
slowly moving older residents
(electrons, He atoms, and dust grains) of the torus,
 may provide sufficient excitation to  cause emission.
Since the He\,{\sc i} emission may come from regions spread across
the torus,
 expansion (or
rotation) of the torus will provide a roughly parabolic emission
line profile provided that the number of contributing
clouds is not small. A velocity shift from the photospheric
velocity will occur if the clouds are asymmetrically
around the torus.

Evidence for excited He atoms along the line of sight to RCB
stars is provided by Clayton, Geballe, \& Bianchi's (2003)
observations of the He\,{\sc i} 10830 \AA\ line in a sample
of RCB stars. In stars observed near maximum light, the 10830 \AA\
line with the exception of V854 Cen is strongly in absorption
with a blueshift of 200--300 km s$^{-1}$. There is a hint of
P Cygni emission with a suggestion that the emission is
stronger for stars observed in decline. Although a detailed
study of He\,{\sc i} excitation is needed, these
observations are consistent with our speculation about the origin of
the the 7065 \AA\ and other lines, all of which must
be much weaker than the 10830 \AA\ line.

\item
When the star is thoroughly obscured by the dust cloud
responsible for the decline, it is seen by starlight
scattered multiple times by 
 dust in the torus. These
dusty clouds  expanding outward  will impose a range of
Doppler shifts on the scattered starlight. The result is
a greatly smeared (veiled) version of the photospheric spectrum is
received by the observer.

\item
In the recovery phase, the sightline to the star is becoming
less opaque. Gas along this sightline may be seen in
absorption against the reduced photospheric continuum, seen
through the new cloud and also be scattering off other
dust clouds. Some of 
this gas is associated with the dust which caused the decline and
some may be in the expanding torus as a result of earlier declines. Dust is
accelerated by radiation pressure and gas-dust collisions
transfer momentum to the gas. This outward moving gas provides
the blue-shifted absorption lines in the Na D and other resonance
lines of abundant atoms and singly-charged ions (e.g., Ca$^+$).
In a very deep decline, the starlight is scattered into the
line of sight from dust primarily off the line of sight to
the star.
The absence of Na D absorption at this
time implies that the high velocity  gas is confined
to a plane and is not spherically distributed about the star.
\end{itemize}

This sketch of the jet-torus model does not address why a RCB
star should have a preferred plane for formation and ejection of
dust and a pair of polar jets. The key to the preferred
plane may possibly be linked to the origin of a RCB
star. A favoured model involves the accretion of a He white
dwarf by a C-O white dwarf. Accretion may  lead to
heating and swelling of the merged envelope, i.e., to the
observed supergiant. One might suppose that C-O white
dwarf is greatly spun up during accretion and, therefore, the
RCB retains a memory of the binary orbital plane. Origins of
the bipolar jets and the formation of dust clouds in the
equatorial plane are left for a future theoretical study.

\section{Acknowledgements}

    We would like to thank Sunetra Giridhar for help with the analysis
of the 1991 and 1992 spectra.
The observations presented here  have been obtained at the
 Cerro Tololo Inter-American
Observatory, National Optical Astronomy Observatories (NOAO), which is operated
by the Association of Universities for Research in Astronomy Inc. (AURA) under
a cooperative agreement with the National Science Foundation, USA. All three 
authors have been Visiting Astronomers at CTIO and would like to thank the 
support staff for their assistance.

\label{lastpage}

\end{document}